\documentclass[aps,prl,onecolumn,superscriptaddress,groupedaddress]{revtex4}
\usepackage{subfig}
\usepackage{graphicx}  % needed for figures
\usepackage{dcolumn}   % needed for some tables
\usepackage{bm}        % for math
\usepackage{amssymb}   % for math
\usepackage{slashed}
\usepackage{graphicx}				% Use pdf, png, jpg, or eps with 	
\usepackage{amsmath}
\usepackage{mathtools}
\usepackage{tikz,pgf}
\usepackage{comment}
\usepackage{asymptote}
\usetikzlibrary{arrows,backgrounds}
\usetikzlibrary{fit,scopes,calc,matrix,positioning,decorations.pathmorphing}
\usepackage[all]{xy}
\usepackage{yfonts}

\newcommand{\ket}[1]{\ensuremath{\left|#1\right\rangle}}
\newcommand{\Bracket}[1]{\ensuremath{\left\langle#1\right\rangle}}

\begin{document}
\title{The Universal Coefficient Theorem, Wormholes, and the Island}
\author{Andrei T. Patrascu}
\address{ELI-NP, Horia Hulubei National Institute for R\&D in Physics and Nuclear Engineering, 30 Reactorului St, Bucharest-Magurele, 077125, Romania}
\begin{abstract}
I analyse the extension of the Ryu Takayanagi formula from the perspective of the universal coefficient theorem and cohomology with different coefficients. Looking at the axioms of cohomology, the so called Eilenberg-Steenrod axioms, the third axiom states that the cohomology of a point is trivial in all orders. If this axiom is abandoned the point may have additional structure. This phenomenon is not unusual in physics, a simple example for that being the spin structure or even the point particle to string transition. I studied the transition between different coefficients in cohomology and hence between different structures of the geometrical point by means of the universal coefficient theorem.  The results may have impact on recent island calculations of the black hole radiation entropy and the Page curve as well as on a potential mapping of higher genus diagrams to genus one diagrams in the topological expansion of string theory. 
\end{abstract}
\maketitle
\section{Introduction}
The Eilenberg Steenrod [1] axioms are the defining ideas behind the classical concept of (co)homology. As we know, cohomology is a tool capable of identifying and counting topological features as well as to classify topological structures of similar type. They are neither unique nor absolute. A major development arose in the study of cohomology by an observation of Grothendieck who, while abandoning the third axiom (of dimension) opened the way to the creation of more generalised cohomologies that could incorporate not only tools of measuring a broader variety of topological features but also could show us how transitions between different types of structures supplemented to the geometrical point can manifest themselves [2]. That Grothendieck's work has not yet been used in understanding quantum phase transitions or dualities like the holographic one remains a mystery. Noticing the great impact of Grothendieck's discovery, I will start describing how his work is linked to understanding the universal coefficient theorem from a physical point of view. Grothendieck created not only the above mentioned extension of cohomology but also the so called "Grothendieck's point of view" which focused on the consideration of classes of objects and the mappings between them as well as the transitions between different classes of parameters describing such geometrical objects as the main tool of understanding their fundamental properties [3]. No more was a single object the focus of geometry, but the whole class of objects and the ways in which it can be linked to its potential deformations. We take this for granted in analysing moduli spaces in string theory nowadays but his work was decades ahead of the work of string theorists and put the foundations of what is known today as algebraic geometry. His work also influenced category theory and categorial logic, among others. 
The universal coefficient theorem [4] basically states that cohomology with a certain type of coefficients can be mapped into (co)homology with a different type of coefficients while enlarging the exact sequence defining it by means of an extension (torsion). The extension plays a role in modifying the algebraic operation that defines the type of transformations performed in the space described by the specific (co)homology when one "forgets" the underlying structure associated to one type of coefficients. This amounts to a change in the integration, leading to new topological terms that found an application in anomaly cancellations [5].  

One of the most calculable situations in which universal coefficient theorems can be applied is the $\mathbb{Z}_{n}$ group which so happens to be at the foundations of the centre symmetry in condensed matter systems. Breaking and restoration of this symmetry lies at the basis of confined-deconfined phase transitions in QCD among others. It is interesting to note that breaking or restoring the $\mathbb{Z}_{n}$ symmetry in the context of cohomology of coefficients amounts to applying the universal coefficient theorem for a transition from a structureless point to a point with a certain structure imposed on it. This expands the associated exact sequence by a term including the extension of this symmetry group. It is also not surprising that such transitions have been analysed in the holographic context as changes in the D-brane structures of the bulk space [6]. This follows a set of steps: first we extend the dimensionality of the system by introducing the bulk space. Then we introduce D-branes which amount to a large (non-perturbative) expansion of the structure of our simple point. Observing that a topological change in the bulk D-brane amounts to a change of phase in our boundary field theory is equivalent with saying that our universal coefficient theorem replaces a simple point structure (a trivial group) with a topologically non-trivial point structure (a $\mathbb{Z}_{n}$ symmetry group). One may hope that at this point the relevance of the universal coefficient theorem becomes clearer both in understanding quantum (potentially holographic) phase transitions as well as in understanding potential extensions of the holographic principle itself. 

\section{Multiple Trace Operators}
A standard anti-de-Sitter space duality can be implemented with a boundary condition corresponding to a double trace operator in the dual field theory [7]. This can link two different conformal field theories which in the large N limit are connected by a Legendre transformation. From a quantum field theoretical perspective one can analytically compute the changes in observables under a renormalisation group flow between fixed points when a small parameter is present and the RG flow is triggered by a double trace deformation of a conformal field theory. A double trace flow connecting  UV and IR points can be defined for a d-dimensional CFT that admits a large $N$ expansion and possesses a single trace operator $\mathcal{O}$ with the conformal dimension $\Delta<d/2$. The double trace deformation of the action is
\begin{equation}
S_{\lambda}=S_{CFT}+\lambda\int d^{d}x \mathcal{O}^{2}
\end{equation}
The unperturbed CFT in the UV flows towards a new CFT in the IR where the operator $\mathcal{O}$ has a dimension $d-\Delta+\mathcal{O}(1/N)$. This corresponds to changing the boundary condition on the bulk scalar field $\phi$ dual to $\mathcal{O}$. The mass of the bulk scalar is given by $m^{2}L_{AdS}^{2}=\Delta(\Delta-d)$ and for $d/2-1<\Delta<d/2$ we have two unitary conditions. One of them $phi\sim z^{\Delta}$ corresponds to the UV CFT while $\phi\sim z^{d-\Delta}$ corresponds to the IR CFT reached because of the double trace perturbation. 
If the bulk contains a traversable wormhole we need some form of interaction between the two CFTs on the boundary [8]. Intuitively enough, this is obtained by turning on a double trace deformation
\begin{equation}
\delta S= \int dt d^{d-1}x h(t,x)\mathcal{O}_{R}(t,x)\mathcal{O}_{L}(-t,x)
\end{equation}
where the boundary scalar operator of dimensions smaller than $d/2$, $\mathcal{O}$ is dual to the scalar field $\phi$ where $h(t,x)$ is a small deformation. It has been seen that in this case a small spherically symmetric perturbation of the stress tensor $T_{\mu\nu}\sim\mathcal{O}(\epsilon)$ results in a traversable wormhole when the ANEC is violated. 
The fact that a traversable wormhole geometry is made possible by double trace insertions is not surprising if one considers how such operators are being introduced. The large-N limit can offer us more perturbative control over interacting systems. The quantum field theories involved in the AdS/CFT duality are seen as matrix large-N field theories having as paragon the non-Abelian Yang Mills theory which is also the basis for the standard model. In this context $N$ represents the number of colours with QCD having $N=3$ while a theory with "any" number of colours would have the limit $N\rightarrow \infty$. There are of course three colour charges carried by quarks and $N^{2}$ gluons exchanging the colours between these sources. These gluons are described by means of a matrix. In the large-N (arbitrary N) limit a mean field is realised by means of a strongly coupled dynamics. This is in contrast with the usual effectively free theory realised in the vector theory models. This mean field theory defines the relation between the boundary and the classical limit in the gravitational bulk. Let us look first at a large-N vector model. As is the case in semi-classical mean field theories like Hartree Fock, controlled by the smallness of $\hbar$, in the large N limit one finds frozen out order parameters. While these are different compared to the semi-classical parameters, the mean field theory is a free theory which receives perturbative corrections under the $1/N$ expansion. Considering for example the Gross-Neveu model in d=1 dimensions or the Nambu-Jona-Lasino model in d=3 dimensions with N species of fermions and four-fermion self interaction we obtain the action
\begin{equation}
S=\int d^{d}x dt(-i\bar{\psi}_{i}(\gamma^{\mu}\partial_{\mu}+m)\psi_{i}+\frac{\lambda}{6}(\bar{\psi_{i}}\psi^{i})^{2})
\end{equation}
where the rotations are generated by $M_{\mu\nu}=\frac{1}{4}[\gamma_{\mu},\gamma_{\nu}]$ with $\{\gamma_{\mu},\gamma_{\nu}\}=2\eta_{\mu\nu}$. If $\lambda$ is small we can use it as a perturbative expansion parameter. If it becomes large for a large number of species one can use the smallness of $1/N$ to perform a stable perturbative expansion. Using a Hubbard-Stratanovich transformation with an order parameter field $\sigma$ the $U(N)$ four fermion theory becomes
\begin{equation}
S=\int d^{d}dt(-i\bar{\psi}_{i}(\gamma^{\mu}\partial_{\mu}+m)\psi^{i}-\frac{3}{2\lambda}\sigma^{2}-\sigma\bar{\psi}_{i}\psi^{i})
\end{equation}
The N-fermionic fields can now be integrated out and we obtain
\begin{equation}
\begin{array}{c}
S=\int d^{d}x dt(-\frac{3}{2\lambda}\sigma^{2}-i\frac{N}{2}Tr ln(-\gamma^{\mu}\partial_{\mu}-m-i\sigma))=\\
\\
=N\int d^{d}x dt(-\frac{3}{2\hat{\lambda}}\sigma^{2}-i\frac{1}{2}Tr ln(-\gamma^{\mu}\partial_{\mu}-m-i\sigma))
\end{array}
\end{equation}
the trace being over the spinor indices and the redefinition of $\lambda$ is such that $N$ becomes an overall factor. In the large $N$ limit with a fixed $\hat{\lambda}$ the theory will be dominated by the saddle points of this action. The order parameter is therefore frozen and becomes a fermion scattering potential. However each saddle point is still a function of the coupling constant $\hat{\lambda}$ and therefore the large $N$ limit can capture non-trivial quantum physics which cannot be seen in a weak coupling or semi-classical approach although one solves saddle point equations. One can analyse the system perturbatively in $1/N$ with a diagramatic structure of a standard weak coupling perturbation theory. In the large $N$ limit one sums up a subset of the diagrams where fermion loops are linked by the auxiliary field but not those where the auxiliary field propagates within a loop. In terms of the original model we approximate each correlation function by its maximal loop decomposition similarly to the RPA semi-classical mean field theory
\begin{equation}
\Bracket{ \bar{\psi}_{i}\psi_{j}\bar{\psi}_{k}\psi_{l} } \sim \Bracket{ \bar{\psi}_{i}\psi_{j}}\Bracket{\bar{\psi}_{k}\psi_{l}}+\mathcal{O}(\frac{1}{N})
\end{equation}
In the case of matrix field theories the mean field theory in the large-N limit is substantially different from the conventional saddle-point approach. In particular the effective theory can remain strongly interacting. Consider a vector $U(1)^{N}$ theory 
\begin{equation}
S=\int d^{d}x dt(-\frac{1}{4g^{2}}F^{i}_{\mu\nu}F_{i}^{\mu\nu})
\end{equation}
One can think of this theory as a Higgsed non-Abelian $U(N)$ theory instead of $N$ species of $U(1)$ symmetries. In the large N limit in the $U(N)$ theory the elementary field in the form of the vector potential $A_{\mu}=A_{\mu}^{a}T_{a}$ spans the generators $T_{a}$ of $U(N)$ with $T_{a}$ the $N\times N$ matrices obeying $[T_{a},T_{b}]=if_{ab}^{c}T_{c}$. The elementary gauge field $A_{\mu}$ therefore is seen as an $N\times N$ matrix rather than an $N$ component vector. The limit in which the rank of the matrix is taken to infinity is fundamentally different from the one for the vector theory. The action for our $U(N)$ theory is then
\begin{equation}
S=\int d^{d}x dt [-\frac{1}{4g^{2}}Tr(F_{\mu\nu}F^{\mu\nu})]
\end{equation}
with the non-linear field strength 
\begin{equation}
F_{\mu\nu}=\partial_{\mu}A_{\nu}-\partial_{\nu}A_{\mu}-i[A_{\mu},A_{\nu}]
\end{equation}
We can expand the gauge field perturbatively in the $N^{2}$ generators $T_{a}$ of the gauge group yielding the standard Yang-Mills Feynman rules. However, 't Hooft [9] considered that we should keep $A_{\mu}$ as an $N\times N$ matrix instead. When the field is a matrix one can track the charge associated to each index separately, leading to the well known double-line notation which allows us to track the power of $N$ of each diagram. Each closed loop gives a power of $N$. Also each Feynman diagram is a tiling of closed loops. The Feynman diagram therefore is represented as a two-dimensional tiled surface with a factor $N^{F}$ where $F$ is the number of faces (i.e. loops). Each Feynman diagram therefore scales as 
\begin{equation}
Diag.=g^{2P-2V_{3}-2V_{4}}N^{F}
\end{equation}
where $P$ is the number of propagators and $V_{i}$ is the number of vertices with $i$ legs. If we consider this a two-dimensional surface then the combination 
\begin{equation}
P-\sum_{i}V_{i}-F=-\chi
\end{equation}
is the Euler characteristic. If we define an effective 't Hooft coupling $\lambda=g^{2}N$ then in the large $N$ limit with $\lambda$ fixed the theory reduces to only planar diagrams. The complete set of planar diagrams is a non-trivial function of the coupling constant and hence the planar simplification here retains much of the physics of the theory. The matrix large-$N$ theory continues to be strongly coupled. However, as opposed to the "classicalisation" of the vector theory, in this case this planar limit "classicalisation" requires gravity. The reorganisation of Feynman diagrams in the form of two-dimensional surfaces of different topologies has interesting consequences. The invariant quantities are traces of products of fields. The matrix generalisation of the $\lambda\phi^{4}$ theory is given by 
\begin{widetext}
\begin{equation}
S=\int dt d^{d}x(Tr(-\partial_{\mu}\bar{\Phi}\partial^{\mu}\Phi-m^{2}\bar{\Phi}\Phi)-\frac{\lambda_{1}}{4!}Tr(\bar{\Phi}\Phi\bar{\Phi}\Phi)-\frac{\lambda_{2}}{4!}Tr(\bar{\Phi}\Phi)Tr(\bar{\Phi}\Phi)-\frac{g^{2}}{4!}Tr([\bar{\Phi},\Phi][\bar{\Phi},\Phi]))
\end{equation}
\end{widetext}
The second to last term is the double trace interaction. It is a subleading contribution to the $1/N$ expansion. If we include the full correlation function hence also the disconnected diagrams we see that the disconnected diagrams scale like $N^{6}$ as compared to the $N^{4}$ scaling of the connected diagrams. The disconnected diagrams dominate the large $N$ limit for any correlation function. This means that in the large $N$ limit correlation functions of gauge invariant operators always factorise 
\begin{equation}
\lim_{N\rightarrow \infty}\Bracket{\mathcal{O}\mathcal{O}}=\Bracket{\mathcal{O}}\Bracket{\mathcal{O}}+...
\end{equation}
and hence large-N gauge invariant operators behave essentially like classical variables in a stronger sense as the full variance also vanishes. This means that the whole ensemble collapsed to a single point. This has been observed by Witten who postulated the existence of a master-field formulation of the theory from where the selection of this configuration is manifest. The dynamical variables in the path integral should be redefined in terms of new degrees of freedom in which the path integral explicitly localises on a single configuration of these fields. Also in the large $N$ limit all expectation values of single trace operators become the evaluation of the operators on this single configuration 
\begin{equation}
\Bracket{\mathcal{O}}=\mathcal{O}(\phi^{cl})
\end{equation}
The large-$N$ limit also implies the cancelling of the mixing between single and double trace operators. Because in the large-$N$ limit the two dimensional surface formed by Feynman diagrams started to look more like the two-dimensional worldsheet of a string the master formulation was expected to be related to string theory. The explicit formulation of such a string-like master equation is only known in certain limiting cases, the most famous (and up to now except some other attempts, the only) such case being AdS/CFT. 
When the 't Hooft coupling $\lambda=g^{2}N$ grows, the tiling formed out of the Feynman diagrams becomes denser and denser. In the case of a $(1+1)$ dimensional gauge theory without dynamical degrees of freedom it has been shown directly that the result was a string theory. However, an explicit formulation of a string-like master equation has not been constructed until the AdS/CFT duality. To understand the string-like nature of a matrix theory in the large N limit one has to consider another notion of duality. 
Consider again the matrix $U(N)$ non-Abelian Yang Mills theory 
\begin{equation}
S=\int d^{d}x dt (-\frac{1}{4g_{YM}}Tr(F_{\mu\nu}F^{\mu\nu}))
\end{equation}
we combine the parameters $N$ and $g_{YM}^{2}$ into $\lambda=g_{YM}^{2}N$. For a large $N$ situation we can use the topological expansion and focus only on the planar diagrams. The planar diagrams are however still a perturbative expansion in the coupling constant $\lambda$. Each planar diagram will have its loop expansion. The string like picture is expected to emerge when $g^{2}_{YM}N$ becomes large and corresponds to the strong coupling regime of the theory. When the diagrams start becoming strings we obtain a non-perturbative domain that can be described in this way. This is the beginning of a string-gauge duality which will (given certain limits) become the AdS/CFT correspondence. It basically means that the most appropriate description of the theory changes as one changes the coupling constant. It is important to note that in this case the duality occurs in the large $N$ regime of the theory. For a small 't Hooft coupling $\lambda$ one has a planar diagram theory while for a large 't Hooft coupling the AdS/CFT duality identifies the dual theory as a quantum gravitational string theory in a curved spacetime with one additional space dimension.

The AdS/CFT correspondence has been discussed in general in the large $N$ limit and the given gauge theory became dual to classical supergravity. In the opposite large $1/N$ limit, the duality is expected to remain valid, however with gravity corrections from perturbative string theory and eventually from M theory. This limit is still unexplored and little is known about the M-theory objects involved. However, there is one aspect of the universal coefficient theorem which may play a fundamental role in understanding such contributions. Indeed, the AdS/CFT duality appeared because the preferred theory for a large $N$ limit in the case in which $\lambda$ was large led us to a classical gravitational description in a bulk space. However, this is based on planar diagrams only. The main question one may ask is what happens when one considers non-planar diagrams and hence one involves a finite $N$ problem? Indeed, the mathematics required to understand such effects is subtle and not commonly employed in physics nowadays. It revolves around what we call the Universal Coefficient Theorem and Cohomology with non-trivial coefficients. The coefficient structure in cohomology is assumed to be trivial and given by the lowest cohomology of the point, all higher cohomologies of the point being nil. Indeed, the cohomology of a point can be considered to be a matter of choice, usually given by $H_{0}(P)=\mathbb{Z}$ or $H_{0}(P)=\mathbb{R}$ with $H_{i}(P)=0$ for all $i>0$. If we abandon this axiom we have several possible situations: first we may consider non-trivial lowest order cohomology of the point, for example involving some parity symmetry $H_{0}(P)=\mathbb{Z}_{n}$, this non-trivial group giving the coefficient structure of the cohomology in general. But this is not all. If we are to abandon this axiom, we may as well abandon it entirely and assume there are non-trivial higher cohomologies associated to our point. In fact, if we look back at the AdS/CFT in the large $N$ limit we will note that as we go towards the non-perturbative regime the fact that a string structure becomes manifest is basically just the fact that our point structure is replaced with that of a string, and a string is expected to be a limit of a brane structure that can have various topological features. Before we go into an analysis of the Universal Coefficient Theorem structure of the finite N contributions I wish to investigate the origin of the AdS/CFT duality in some more detail. As we know there are two types of strings, closed and open. While for a particle the dynamics is given by a worldline action 
\begin{widetext}
\begin{equation}
S_{particle}=\frac{1}{2}\int_{\tau_{0}}^{\tau_{1}} d\tau \sqrt{-h(\tau)}[-h^{-1}(\tau)g_{\mu\nu}(x(\tau))\frac{dx^{\mu}(\tau)}{d\tau}\frac{dx^{\nu}(\tau)}{d\tau}-m^{2}-iqA_{\mu}(x(\tau))\frac{dx^{\mu}(\tau)}{d\tau}]
\end{equation}
\end{widetext}
where $g_{\mu\nu}$ is the spacetime metric, $m$ is the mass, $A_{\mu}$ is the background vector potential 
and $h(\tau)$ the worldline metric guaranteeing invariance under proper time transformations $\tau\rightarrow \tau'(\tau)$, the dynamics of a string is derived from a worldsheet action 
\begin{widetext}
\begin{equation}
S_{open}=\frac{1}{l_{s}^{2}}\int d\tau\int_{0}^{\pi}d\sigma\sqrt{-h(\tau)}[-G_{\mu\nu}(X(\sigma,\tau))h^{\alpha\beta}\partial_{\alpha}X^{\mu}(\sigma,\tau)\partial_{\beta}X^{\nu}(\sigma,\tau)]-\oint_{\sigma=0,\pi}d\tau A_{\mu}(X)\partial_{\tau}X^{\mu}
\end{equation}
\end{widetext}
and
\begin{widetext}
\begin{equation}
S_{closed}=\frac{1}{l_{s}}\int d\tau \int_{0}^{2\pi}d\sigma\sqrt{-h(\tau)}[-G_{\mu\nu}(X(\sigma,\tau))h^{\alpha\beta}\partial_{\alpha}X^{\mu}(\sigma,\tau)\partial_{\beta}X^{\nu}(\sigma,\tau)]
\end{equation}
\end{widetext}
where $G_{\mu\nu}$ is the background spacetime metric, $h^{\alpha\beta}$ is the worldsheet metric with determinant $h$ and string theory provides its own background fields. The lowest energy fluctuations surviving the limit $l_{s}\rightarrow 0$ in the case of the open string are vector like gauge fields and in the case of the closed string are gravitons serving as non-linear factors of the kinetic term. Open strings must co-exist with closed strings. A one loop diagram of an open string is equal to a tree level closed string diagram after a reparametrisation of the local worldsheet coordinates $\tau$ and $\sigma$. This connection is called the open-closed string duality. This appears to be the origin of the AdS/CFT duality itself. In the open string situation for each spatial direction there can be implemented a boundary condition corresponding to free moving endpoints having fixed velocity $\partial_{\tau}X^{i}(\sigma=0,\pi)=0$. Another option is to fix the endpoints at a specific location $X^{i}=const.$ The last type of conditions are the Dirichlet conditions and the hyperplane spanned by the condition $X^{i}=const.$ is the Dirichlet brane. The energy of this brane scales as $e^{-1/g_{s}}$ where $g_{s}$ is the string coupling constant. D-branes are seen as non-perturbative solitons of string theory. Solitons are localised finite-energy solutions to the equations of motion. They always break translation invariance. The small excitations around the soliton contain a zero mode corresponding to a translation of the soliton centre. This zero mode is described as quantum mechanics on the worldline of the soliton. For an open string there are two types of low energy excitations with Dirichlet boundary conditions: gauge fields localised on the brane, and the zero modes of the soliton localised on the worldvolume of the brane. The low energy effective action of a D-brane (a $p+1$ dimensional brane) is
\begin{equation}
S_{le}=\frac{1}{g_{s}}\int d^{p}x dt (-\frac{1}{4}F_{\mu\nu}F^{\mu\nu}-\frac{1}{2}\partial_{\mu}\phi^{i}\partial^{\mu}\phi_{i})
\end{equation}
where $\phi^{i}$ are the translational zero modes with $i=1,...,d-p$ describing the directions transverse to the $Dp$-brane. In string theory we need the total number of spacetime dimensions to be 10 and hence here $d=9$. The theory is also supersymmetric and hence we have 16 supersymmetries. For $p=2n-1$, $2n$ dimensions we also have $16/2^{n}$ species of Majorana or Weyl fermions. In the low energy effective action for a D3-brane we have four species of Majorana fermions. We have a symmetry that connects four species of fermions which become exactly $SO(6)=SU(4)$. This is the resulting rotational symmetry around the D3-brane. Scalars and fermions rotate under the same symmetry. The action can be represented via the anti-symmetric representation of $SU(4)$
\begin{equation}
S_{D3}=\frac{1}{g_{s}}\int d^{3}x dt (-\frac{1}{4}F^{\mu\nu}F_{\mu\nu}-\frac{1}{2}\partial_{\mu}\phi^{AB}\partial^{\mu}\phi_{AB}-i\bar{\psi}^{A}\gamma^{\mu}\partial_{\mu}\psi_{A})
\end{equation}
with $A=1,...,4$ and $\phi_{AB}=-\phi_{BA}$. Let us now consider a set of N Dp-branes near each other. Each brane has its own gauge fields. We can see strings as thin extensions of the Dp-brane geometry such that they appear to be stretching between each pair of brains $i$ and $j$. In the proper limit the lowest excitations of those strings has a mass depending on the relative positions of the D-branes in the mutually orthogonal directions $m_{ij}=(r_{i}-r_{j})/l_{s}^{2}$. If the D-branes are located exactly one on top of another then these excitations will be massless. This mode however is also a vector. These vector like modes enhance the gauge group from $U(1)^{N}$ to $U(N)$. On the other side the relative distance between the branes is also a zero mode and this must be combined with the original translational zero modes producing our matrix $N\times N$ field theory with a potential reproducing that of vector fields. The assumption that the relative distance between the branes is a zero mode is only a lowest order approximation. The zero mode is lifted at first order for most quantum field theories and hence solitons interact with each other. We call mutually BPS those D-branes in which by means of supersymmetry we can show that the relative distance is an exact quantum zero mode. Changes in the labelling of the Dp-branes correspond to rotations in the gauge group $U(N)$ and hence the scalar zero modes are charged under the $U(N)$ symmetry. We obtain hence the bosonic part of the low energy effective action as
\begin{widetext}
\begin{equation}
S_{le}=\frac{1}{g_{s}}\int d^{p}x dt [-\frac{1}{4}Tr(F_{\mu\nu}F^{\mu\nu})-\frac{1}{2}Tr(D_{\mu}\phi^{AB}D^{\mu}\phi_{AB})-\frac{1}{4}Tr([\phi^{AB},\phi^{CD}][\phi_{AB},\phi_{CD}])]
\end{equation}
\end{widetext}
where $D_{\mu}\phi_{AB}=\partial_{\mu}\phi_{AB}+i[A_{\mu},\phi_{AB}]$. If we are to introduce also the $16/2^{n}$ species of $N\times N$ charged fermions with a Yukawa interaction to the scalars from supersymmetric strings we can write for a D3 brane example the action as
\begin{widetext}
\begin{equation}
S_{le}=\frac{1}{g_{s}}\int d^{p}x dt [-\frac{1}{4}Tr(F_{\mu\nu}F^{\mu\nu})-\frac{1}{2}Tr(D_{\mu}\phi^{AB}D^{\mu}\phi_{AB})-\frac{1}{4}Tr([\phi^{AB},\phi^{CD}][\phi_{AB},\phi_{CD}])-i Tr \bar{\psi}^{a}\gamma^{\mu}D_{\mu}\psi_{A}-Tr\bar{\psi}^{A}[\phi_{AB},\psi^{B}]]
\end{equation}
\end{widetext}
We have 16 supersymmetries that transform all fields into each other and form a single representation of $\mathcal{N}=4, p=3+1$ supersymmetry where $\mathcal{N}$ represents the number of independent spinor charges. The theory is an interacting theory with an exact electric-magnetic duality. The dimensionless coupling constant does not flow and hence the theory has no intrinsic scale i.e. it is a super-Yang-Mills conformal field theory. On the other side the closed string point of view includes gravitons. D-branes have a finite energy and they must have a representation on the closed string side. D-branes are also charged under one of the additional set of bosonic closed-string fields, the Ramond-Ramond potentials with their charge equaling their mass. Such a charge-mass relation is found in the Reissner-Nordstrom black hole solutions. In this case having $(p+1)$-dimensional membranes we have actually black branes. As we analyse the N coincident D3-branes we can see them as $(3+1)$-dimensional black hole solitons in a $(9+1)$ dimensional space-time. The extremal solution is the metric
\begin{widetext}
\begin{equation}
\begin{array}{c}
ds^{2}=H(r)^{-1/2}(-dt^{2}+dx_{1}^{2}+dx_{2}^{2}+dx_{3}^{2})+H^{1/2}(r)(dr^{2}+r^{2}d\Omega^{2}_{S^{5}})\\
\\
H(r)=1+\frac{4\pi g_{s}N l^{4}_{s}}{r^{4}}
\\
\end{array}
\end{equation}
\end{widetext}
where $d\Omega^{2}_{S^{5}}=d\theta^{2}+sin^{2}\theta d\Omega^{2}_{S^{4}}$ is the metric on the five dimensional unit sphere, $l_{s}$ is the $(9+1)$ dimensional Planck length and $g_{s}$ is the string coupling constant. With them we can also derive Einstein's equations in $(9+1)$ dimensions from string theory as
\begin{equation}
R_{\mu\nu}-\frac{1}{2}g_{\mu\nu}R=\frac{1}{2\pi}g_{s}^{2}(2\pi l_{s})^{8}T_{\mu\nu}
\end{equation}
We also have a generalisation of the Maxwell field strength in the form of a rank five anti-symmetric form
\begin{equation}
F_{txyzr}+\frac{1}{5!}\epsilon_{txyzr\alpha\beta\gamma\delta\zeta}F^{\alpha\beta\gamma\delta\zeta}=H^{-2}(r)\frac{16\pi g_{s} N l^{4}_{s}}{r^{5}}
\end{equation}
Open-closed string duality means that a closed string theory in the background of the above black brane solution represents the same physics as the full open and closed string theory with a low energy effective action given by $\mathcal{N}=4$ super-Yang-Mills theory. Decoupling parts of the degrees of freedom on both sides leads to the AdS/CFT duality. What happens is that if we take $l_{s}\rightarrow 0$ on the open string side, since the gravitational coupling constant generating the interaction between open and closed strings is dimensionful, the open string degrees of freedom decouple from the gravitational closed string sector and become their low energy domain, namely a pure $\mathcal{N}=4$ super-Yang-Mills theory. Maldacena demanded that the Higgs mass $m_{H}\sim r/l_{s}^{2}$ stays fixed as we take on the closed string side the limits $\l_{s}\rightarrow 0$  and $r\rightarrow 0$ simultaneously. This guarantees that the same limit is taken on the black brane closed string side. Then the full theory splits into two parts that will become asymptotically separated: closed strings in flat space and closed strings in the near horizon limit. the near horizon metric is
\begin{equation}
ds^{2}=\frac{r^{2}}{L^{2}}(-dt^{2}+dx_{1}^{2}+dx_{2}^{2}+dx_{3}^{2})+\frac{L^{2}}{r^{2}}(dr^{2}+r^{2}d\Omega_{S^{5}}^{2})
\end{equation}
which represents an anti-de-Sitter space times a five-sphere. In their respective limits both the open and closed string theories result in a decoupled flat-space closed string sector times a different sector. Since the original open-closed string theories were dual these two sectors are equivalent, but that also means that the open string sector of $\mathcal{N}=4$ super-Yang-Mills should be equivalent to closed strings living in the near horizon limit namely $AdS_{5}\times S^{5}$ explicitly containing gravity. This is the short story of the AdS/CFT duality. 
Looking at the super Yang-Mills theories by means of a holographic description in terms of IIB string theory on an anti-de-Sitter background we can note that the effective supergravity of string theory reaches the strong coupling limit of the conformal field theory in the large $N$ limit. String loop corrections correspond to the expansion in inverse powers of the 't Hooft effective coupling $g_{YM}^{2}N$. Correlation functions in the $\mathcal{N}=4$ super-Yang-Mills theory at large $N$ are determined through the standard relation 
\begin{equation}
\prod_{j=1}^{k}(\frac{\delta}{\delta \phi_{0,j}(z_{j})})\left.e^{i S_{sugra}[\phi(\phi_{0})]}\right|_{\phi_{0,j}=0}=\Bracket{\prod_{j=1}^{k}\mathcal{O}^{j}(z_{j})}_{CFT}
\end{equation}
where the action $S_{sugra}$ refers to the bulk action of supergravity as a functional of the boundary values of the fields. $\phi_{0,j}$ and $\mathcal{O}$ are the composite operators of the conformal Yang Mills theory. The boundary values of the supergravity source act as sources of these operators. 
A naive calculation of the correlations of composite operators is generally divergent. This problem is solved by a proper regularization of the short distance singularities. One method of rendering such correlations finite was to compute the supergravity Green's functions at points very near the AdS boundary and to consider this as an IR cut-off for the gravity theory that acts like an UV regulator in the CFT. Counterterms with the corresponding scale are introduced rendering a renormalisation of the CFT through the AdS boundary theory. This would violate conformal invariance but one did not consider the CFT itself but instead a perturbed CFT by the conformal operators 
\begin{equation}
S_{\mathcal{N}=4 SYM}\rightarrow S_{\mathcal{N}=4 SYM}+\int d^{4}x \phi_{0,j}(x)\mathcal{O}^{j}(x)
\end{equation}
The source is the boundary value of the supergravity field. Conformal symmetry however protects the dimensions of chiral primary operators and their descendants. In the case of operators whose sources are elementary supergravity fields, the introduction of a regulator at intermediate steps does not affect the final answer if the operator insertions are kept at separated points. If however we introduce multiple operators the correlation functions may diverge at short distances and that would require additional counterterms. Insertions of operators in the SYM Green's functions produce counterterms of the form of products of conformal operators at the same point. These composite operators look like $:Tr F^{2}(x) Tr F^{2}(x): $ and are dual to multiparticle supergravity states. Such multiparticle states are necessary in the AdS/CFT correspondence. When evaluating three and four point functions explicit contact contributions appear. Their k-space expressions are divergent but they produce logarithms in the kinematic invariants and contain cuts. A four point function will contain a contact term as in 
\begin{widetext}
\begin{equation}
\Bracket{\mathcal{O}(x_{1})\mathcal{O}(x_{2})\mathcal{O}(x_{3})\mathcal{O}(x_{4})}=N^{2}\delta^{(d)}(x_{1}-x_{2})\frac{1}{|x_{1}-x_{3}|^{p}}\delta^{(d)}(x_{3}-x_{4})+...
\end{equation}
\end{widetext}
to which we add products of multiple delta functions. The coincident points relate to the regulating procedure of the divergences and the appearance of contact terms in the calculation of correlation functions. The contact contributions to the multi-point correlation functions can be calculated using certain limits of the bulk-bulk and bulk-boundary kernels. Taking as an example the dilaton-axion sector of the IIB supergravity on $AdS_{5}\times S_{5}$ given by 
\begin{equation}
S=\frac{1}{2\kappa^{2}}\int d^{5}x \sqrt{g(x)}(-R+12/A^{2})+g^{\mu\nu}[\partial_{\mu}\phi\partial_{\nu}\phi+e^{2\phi}\partial_{\mu}C\partial_{\nu}C]
\end{equation}
with the background metric on $AdS_{5}$ given by the half-space Poincare metric
\begin{equation}
\begin{array}{cc}
ds^{2}=\frac{A^{2}}{x_{0}^{2}}(dx^{2}_{0}+dx^{i}dx^{j}\eta_{ij}), & i=1,...,d
\end{array}
\end{equation}
with $x_{0}\geq 0$.
It can be noted that the dilaton-axion coupling contains derivatives. For convenience the AdS radius $A^{2}$ can be set to one. For massive scalars the bulk-bulk correlator is 
\begin{equation}
\begin{array}{c}
\Bracket{\phi(x)\phi(y)}=G(x,y)=\\
\\
=(x_{0}y_{0})^{d/2}\int_{0}^{\infty}\lambda d\lambda\int\frac{d^{d}k}{(2\pi)^{d}}e^{i k\cdot(x-y)}\frac{J_{\nu}(\lambda x_{0})J_{\nu}(\lambda y_{0})}{\lambda^{2}+k^{2}-i\epsilon}
\end{array}
\end{equation}
where $k\cdot x=\sum_{i=1}^{d}k_{i}x^{i}$ with $x$ being a Minkowski four-vector on the boundary of AdS. The bulk-boundary kernel results from the $y_{0}\rightarrow 0$ limit
\begin{equation}
\Delta(y,z)=\lim_{y_{0}\rightarrow 0}\frac{1}{y_{0}^{d/2+\nu-1}}\partial_{y_{0}}G(y,z)
\end{equation}
$y_{0}^{-\nu}$ corrects for the asymptotic behaviour of the Green's function. The asymptotic behaviour of the Bessel functions at $z\rightarrow 0$ is
\begin{equation}
\begin{array}{cc}
J_{\nu}(z)=\frac{1}{\Gamma(1+\nu)}(\frac{z}{2})^{\nu}+... , & K_{\nu}(z)=\frac{\Gamma{\nu}}{2}(\frac{2}{z})^{\nu}+...
\end{array}
\end{equation}
and at $z\rightarrow \infty$ we have
\begin{equation}
\begin{array}{cc}
J_{\nu}(z)=(\frac{1}{2\pi z})^{1/2} cos(z+\pi/2)+..., & K_{\nu}(z)=(\frac{\pi}{2z})^{1/2}e^{-z}+...
\end{array}
\end{equation}
Therefore in the small $z$ limit we find
\begin{equation}
\begin{array}{c}
\Delta(y,z)=(z_{0})^{d/2}\int_{0}^{\infty}\lambda d\lambda\int\frac{d^{d}k}{(2\pi)^{d}}e^{i k \cdot (z-y)}\frac{2}{\Gamma(\nu)}(\frac{\lambda}{2})^{\nu}\frac{J_{\nu}(\lambda z_{0})}{\lambda^{2}+k^{2}-i\epsilon}=\\
\\
\frac{2}{\Gamma(d/2)}\int\frac{d^{d}k}{(2\pi)^{2}}(\frac{|k|\cdot z_{0}}{2})^{d/2} K_{\nu}(|k|\cdot z_{0})e^{i k(x-z)}
\end{array}
\end{equation}
In general $\nu=\sqrt{m^{2}+d^{2}/4}$ but for massless fields $\nu=d/2$ and we obtain after integrating over the Fourier modes
\begin{equation}
\Delta(y,z)=\frac{\Gamma(d)}{\pi^{d/2}\Gamma(d/2)}(\frac{z_{0}}{z_{0}^{2}+(y-z)^{2}})^{d}
\end{equation}
The derivative of the bulk-boundary kernel 
\begin{equation}
\mathcal{F}(y,z)=\lim_{z_{0}\rightarrow 0}\frac{1}{z_{0}^{d-1}}\partial_{z_{0}}\Delta(y,z)
\end{equation}
and we obtain 
\begin{widetext}
\begin{equation}
\mathcal{F}(y,z)=\lim_{z_{0}\rightarrow 0}\frac{\Gamma(d+1)}{\pi^{d/2}\Gamma(d/2)}\{-\frac{z_{0}^{2}}{[z_{0}^{2}+(y-z)^{2}]^{d+1}}+\frac{(y-z)^{2}}{[z_{0}^{2}+(y-z)^{2}]^{d+1}}\}
\end{equation}
\end{widetext}
with the limits
\begin{equation}
\begin{array}{cc}
y-z\neq 0, & \mathcal{F}(y,z)=\frac{\Gamma(d+1)}{\pi^{d/2}\Gamma(d/2)}\frac{1}{(y-z)^{2d}}\\
\\
y-z=0, & \mathcal{F}(y,z)=-\frac{\Gamma(d+1)}{\pi^{d/2}\Gamma(d/2)}\frac{1}{y_{0}^{2d}}
\\
\end{array}
\end{equation}
The $z_{0}\rightarrow 0$ form of the derivative of the bulk boundary kernel is 
\begin{equation}
\lim_{z_{0}\rightarrow 0}\frac{1}{z_{0}^{d-1}}\partial_{z_{0}}\Delta(y,z)=\frac{\Gamma(d+1)}{\pi^{d/2}\Gamma(d/2)}\{\frac{1}{(y-z)^{2d}}-\frac{1}{\mu^{d}}\delta(y-z)\}
\end{equation}
$\mu$ being the infinitesimal distance from the boundary at $z_{0}=0$.

\section{The Universal Coefficient Theorem}
The last concept required for the following parts of this paper is the universal coefficient theorem. As stated at the very beginning in the introduction, (co)homology with coefficients is capable of calculating topological invariants in cases in which the fine graining of the manifold reveals or hides some non-trivial structure. This is particularly important in the case in which we have to move between various limits reaching to certain gauge/string dualities. Indeed, it seems that a general gauge/string duality should be encoded in the link between (co)homology with different coefficients. This link can be formalised in the language of homological algebra as the universal coefficient theorem. Having this in mind one may ask what is the analogue of the renormalisation group equations in the case in which the fine grain structure is topologically non-trivial? What changes when by changing the scale we need to blur out different microscopic topologies? This will prove to be of great importance in understanding string theory's topological genus expansion. In order to properly construct the universal coefficient theorem we need some auxiliary objects, particularly the categorial definition of tensor products, adjoint functors and the $Hom$ group as well as $Tor$ and $Ext$ (the torsion and the extension). Let us start with the tensor product. Let $A$ and $B$ be two modules over a commutative ring $R$. The tensor product of $A$ and $B$ is the $R$-module $A\otimes_{R}B$ defined as the quotient 
\begin{equation}
\frac{F(A\times B)}{R(A\times B)}
\end{equation}
where $F(A\times B)$ is the free $R$-module with basis $A\times B$ and $R(A\times B)$ the submodule generated by 
\begin{equation}
\begin{array}{c}
(a_{1}+a_{2},b)-(a_{1},b)-(a_{2},b)\\
(a,b_{1}+b_{2})-(a,b_{2})-(a,b_{2})\\
r(a,b)-(ra,b)\\
r(a,b)-(a,rb)\\
\end{array}
\end{equation}
The image of a basis element $(a,b)$ can be written in $A\otimes_{R}B$ as $a\otimes b$. We also have the relations
\begin{equation}
\begin{array}{c}
(a_{1}+a_{2})\otimes b=a_{1}\otimes b+a_{2}\otimes b\\
a\otimes(b_{1}+b_{2})=a\otimes b_{1}+a\otimes b_{2}\\
(ra\otimes b)=r(a\otimes b)=(a\otimes rb)\\
\end{array}
\end{equation}
That means we can say that $A\otimes_{R}B$ is the largest $R$-module generated by the set of symbols $\{a\otimes b\}_{a\in A,b\in B}$ satisfying the above product relations. Any element of $A\otimes B$ can be represented as a finite sum $\sum_{i=1}^{n}a_{i}\otimes b_{i}$ with the mention that it may not be possible to take $n=1$. Of course the sum representation is not necessarily unique. 
An $R$-bilinear map $\beta:A\times B\rightarrow C$ is the same as an element of $Hom_{R}(A,Hom_{R}(B,C))$. By the adjoint property of tensor products there exists an isomorphism of $R$-modules of the form
\begin{equation}
Hom_{R}(A\otimes_{R}B,C)\cong Hom_{R}(A,Hom_{R}(B,C))
\end{equation}
natural in $A, B, C$ given by $\phi\leftrightarrow(a\rightarrow(b\rightarrow\phi(a\otimes b)))$. To introduce $Hom$, for an $R$-module $A$ we can define $A^{*}=Hom_{R}(A,R)$. This is often called the dual of $A$. Given an $R$-module map $f:A\rightarrow B$ the dual map $f^{*}:B^{*}\rightarrow A^{*}$ is defined by $f^{*}(\phi)=\phi\circ f$. The operation of taking duals therefore defines a contravariant functor from the category if $R$-modules to itself. For $R$-modules $A$ and $M$, $Hom_{R}(A,M)$ is the $R$-module of homomorphisms from $A$ to $M$. It is contravariant in its first variable and covariant in its second variable. For an $R$-map $f:A\rightarrow B$ we have 
\begin{equation}
Hom_{R}(f,M):Hom_{R}(B,M)\rightarrow Hom_{R}(A,M)
\end{equation}
defined by $\phi\rightarrow \phi\circ f$. Then $f^{*}$ means $Hom_{R}(f,M)$.
The next concepts important for introducing the universal coefficient theorem are $Ext$ and $Tor$. Let there be the $R$-module $R$-Mod and the functor that takes a short exact sequence to a short exact sequence $F:R$-Mod$\rightarrow R$-Mod. A covariant functor is right-exact if $F(A)\rightarrow F(B)\rightarrow F(C)\rightarrow 0$ is a short exact sequence. Similarly a contravariant functor is right-exact if $F(C)\rightarrow F(B)\rightarrow F(A)\rightarrow 0$ is a short exact sequence. The functors defined above: tensor product and homomorphism $*\otimes_{R} M$, $Hom_{R}(M,*)$, and $Hom_{R}(*,M)$ are in general not exact. It is interesting to find natural functors which measure the failure of another functor to preserve short exact sequences. 
  One may try to take for $*\otimes_{R}$ the  kernel  of $A\otimes M\rightarrow B\otimes M$ as  the  value  of  this  functor.   However,  this does  not  behave  nicely  with  respect  to  morphisms.   To  construct  these  functors  the only things we will use are the left/right exactness properties and the observation that for any module $M$ there is a surjective map from a free module to $M$.  Exactness is a very important property.  It essentially means that the objects and morphisms in the sequence are arranged such that the image of one morphism is the kernel of the next.  When we speak about short exact sequences we also have the first map being an injection and the second a surjection. 
There exist two functors. First 
\begin{equation}
Tor_{n}^{R}:R_{Mod}\times R_{Mod}\rightarrow R_{Mod}
\end{equation}
for all $n=0,1,2,...$, with $(M_{1},M_{2})\rightarrow Tor_{n}^{R}(M_{1},M_{2})$ covariant in $M_{1}$ and $M_{2}$ satisfying the following axioms:
\begin{itemize}
\item $Tor_{0}^{R}(M_{1},M_{2})=M_{1}\otimes_{R}M_{2}$\\
\item If $0\rightarrow A \rightarrow B\rightarrow C\rightarrow 0$ is any short exact sequence of $R_{Mod}$ and $M$ is any $R_{Mod}$ then there is a natural long exact sequence 
\begin{equation}
\begin{array}{c}
...\rightarrow Tor_{n}^{R}(A,M)\rightarrow Tor_{n}^{R}(B,M)\rightarrow Tor_{n}^{R}(C,M)\rightarrow Tor_{n-1}^{R}(A,M)\rightarrow ...\\
...\rightarrow Tor_{1}^{R}(C,M)\rightarrow A\otimes_{R}M\rightarrow B\otimes_{R}M\rightarrow C\otimes_{R}M\rightarrow 0\\
\end{array}
\end{equation}
\item $Tor_{n}^{R}(F,M)=0$ if $F$ is a free module and $n>0$
\end{itemize}
The functor $Tor_{n}^{R}(*,M)$ is called the $n^{th}$ derived functor of the functor $*\otimes_{R}M$.
Then we have 

\begin{equation}
Ext_{R}^{n}:R_{Mod}\times R_{Mod}\rightarrow R_{Mod}
\end{equation}
for all $n=0,1,2,...$, with $(M_{1},M_{2})\rightarrow Ext_{R}^{n}(M_{1},M_{2})$ contravariant in $M_{1}$ and covariant in $M_{2}$ satisfying the following axioms:
\begin{itemize}
\item $Ext_{R}^{0}(M_{1},M_{2})=Hom_{R}(M_{1}, M_{2})$\\
\item If $0\rightarrow A \rightarrow B\rightarrow C\rightarrow 0$ is any short exact sequence of $R_{Mod}$ and $M$ is any $R_{Mod}$ then there is a natural long exact sequence 
\begin{equation}
\begin{array}{c}
...\rightarrow Hom_{R}(C,M)\rightarrow Hom_{R}(B,M)\rightarrow Hom_{R}(A,M)\rightarrow Ext_{R}^{1}(C,M)\rightarrow ...\\
...\rightarrow Ext_{R}^{q}(B,M)\rightarrow Ext_{R}^{q}(A,M)\rightarrow Tor_{R}^{q+1}(C,M)\rightarrow...\\
\end{array}
\end{equation}
\item $Ext_{R}^{n}(F,M)=0$ if $F$ is a free module and $n>0$
\end{itemize}
The functor $Ext_{R}^{n}(*,M)$ is called the $n^{th}$ derived functor of the functor $Hom_{R}(*,M)$.
Any functors satisfying the above axioms are isomorphic and these axioms generally characterise the functors $Tor$ and $Ext$. 
Let me now introduce the universal coefficient theorem. Let $(C_{*},\partial)$ be a chain complex over a ring $R$. Then there is an evaluation map 
\begin{equation}
\begin{array}{c}
Hom_{R}(C_{q},M)\times C_{q}\rightarrow M\\
(f,z)\rightarrow f(z)\\
\end{array}
\end{equation}
This pairing passes to the Kroneker pairing 
\begin{equation}
<,>:H^{q}(C_{*};M)\times H_{q}(C_{*})\rightarrow M
\end{equation}
of cohomology with homology. This pairing is bilinear and its adjoint is a homomorphism 
\begin{equation}
H^{q}(C_{*},M)\rightarrow Hom(H_{q}(C_{*});M)
\end{equation}
and this does not need to be an isomorphism. It is not trivial to understand the kernel and co-kernel of this map. This is also the main reason why the large $N$ limit in the genus expansion is usually focused on the planar diagrams but when this limit is lifted it is not trivial to take into account the higher genus topological terms. This is also the main issue in relating string theories to gauge theories in more general cases aside of AdS/CFT. Universal coefficient theorems provide a measure of how such an adjoint fails to be an isomorphism and the discrepancy is given precisely the these $Ext^{q}$ and $Tor_{q}$ functors. The answer is in general not simple. There exists a complete answer to such a question when $R$ is a principal ideal domain and $C_{*}$ is a free chain complex. In this case $H^{q}(C_{*},M)\rightarrow Hom(H_{q}(C_{*});M)$ is surjective with kernel $Ext(H_{q-1}(C_{*}),M)$. This will cover the topological situation when the coefficients are integers or belong to a field since the singular and cellular complexes of a space are free. It is interesting to analyse this looking at the order of the cohomology/homology. The cohomology of order $q$ maps to a homomorphism of homology of order $q$ which is equivalent to saying that the homology group of order $q$ in a classical example would represent a finitely generated abelian group with the generators organised to represent the dimensions of the topologically non-trivial elements. It is therefore interesting albeit not terribly surprising in the context of AdS/CFT to note that the obstruction to the map being an isomorphism is the kernel $Ext(H_{q-1},M)$ which would correspond to a homology group with generators corresponding to topologically non-trivial elements in one dimension lower. 
The universal coefficient theorem can therefore be formulated as follows. Let $R$ be a principal ideal domain. Suppose that $M$ is a module over $R$ and $(C_{*},\partial)$ is a free chain complex over $R$ (each $C_{q}$ is a free $R$-module.) Then the sequence 
\begin{equation}
0\rightarrow Ext_{R}(H_{q-1}(C_{*}),M)\rightarrow H^{q}(C_{*};M)\rightarrow Hom(H_{q}(C_{*}),M)\rightarrow 0
\end{equation}
is exact and natural with respect to chain maps of free chain complexes. Moreover, the sequence splits albeit not naturally. Similarly for homology we have equivalently 
\begin{equation}
0\rightarrow H_{q}(C_{*})\otimes M\rightarrow H_{q}(C_{*}\otimes M)\rightarrow Tor_{1}^{R}(H_{q-1}(C_{*}),M)\rightarrow 0
\end{equation}

%%%

\section{Topology of replica trick manifolds}
The structure of entanglement in the non-gravitational (usually CFT-large-N) system leads to the emergence of gravitational physics via AdS/CFT. It has been shown that areas of extremal surfaces in the dual spacetime describe the entanglement entropy for subsystems at leading order in $N$. Arbitrary first order perturbations of the vacuum state lead to entanglement entropies that can always be described by extremal surfaces in some first order perturbation of the AdS spacetime. By making use of the entanglement first law the perturbed spacetime geometry satisfies the linearised Einstein's equations. In the case of second order perturbations if one adds local sources to the Euclidean path integral describing the CFT vacuum state, the entanglement entropies up to second order perturbation in the sources defining that state can be described geometrically by a second order perturbation to AdS. The structure of the CFT entanglement in this case implies that these perturbations must satisfy local gravitational equations including now non-linear contributions. Such non-linearities emerge from the relation between the CFT entanglement and the one-point functions of local operators. The states defined by single trace sources in the Euclidean path integral correspond to coherent states of the bulk fields. In general there exist bulk gravitational theories that have a good semiclassical description but have a more general bulk entanglement. The Euclidean path integral in this case will involve sources for non-local double trace and general multi-trace operators. The gravity-side entanglement is expected to be given by the quantum Ryu-Takayanagi formula 
\begin{equation}
\Delta S_{A}^{CFT}=\frac{1}{4G}\Delta Area(\tilde{A})+\Delta S_{\Sigma}^{bulk}
\end{equation}
where $\tilde{A}$ is a bulk surface homogenous to $A$ which extremises the expression while the second term is the vacuum subtracted bulk entanglement entropy of the bulk fields in the region $\Sigma$ bounded by this surface. This last term is important for the description of the CFT entanglement for non-coherent bulk states. 
It has been shown that for any state created by general nonlocal sources
\begin{equation}
\int dx_{1}...dx_{n} \lambda_{\alpha_{1}...\alpha_{n}}^{(n)}(x_{1},...,x_{n})\mathcal{O}_{\alpha_{1}}(x_{1})...\mathcal{O}_{\alpha_{n}}(x_{n})
\end{equation}
the CFT entanglement entropies for arbitrary spatial regions at leading order in $1/N$ are the same as for another state with only a single trace source $\lambda_{eff}(x)$ determined in terms of the sources $\lambda^{(n)}$ by means of a self-consistency equation [10]. The effective single trace source depends non-trivially on all the multi-trace sources in general but vanishes if the single trace sources are set to zero. The double and higher-trace sources on their own can change the quantum state and the entanglement structure of the bulk fields but do not cause any backreaction. When however the single-trace sources are turned on, the final backreacted geometry depends on which multi-trace sources are already present. A composite effective single trace source is equivalent to a description where the effects of local multi-trace sources are captured by a modified boundary condition in AdS/CFT. Bulk quantum effects are being described by non-analyticities. In the sub-leading orders in the  $1/N$ expansion there are contributions 
 to the CFT entanglement described by bulk quantum effects given by the quantum Ryu-Takayanagi formula. Contributions to the CFT entanglement entropy at sub-leading orders are fundamentally different from those of the $O(N^{2})$ entropies. When we use a perturbative expansion in the sources $\lambda^{(n)}$, we may obtain divergences at specific orders in the sources if the support of the sources goes beyond a certain strip of analyticity in Euclidean time. Such divergences have been shown to be an artefact of the full expression of the entropies which have non-analytic contributions. Such non-analytic contributions involves a dual gravitational description with non-geometric contributions like, for example, bulk entanglement. 
 These contributions can be described precisely as an effect of the Ext and Tor groups in the universal coefficient theorem and are a result of the need to change the underlying point space topology to one that is non-trivial. This suggests a fine grained entanglement structure formed out of microscopical wormhole geometries spreading at very short scales [11]. Of course such topologies can be avoided by choosing the coefficient structure that makes them invisible at a cost of generating divergences as the ones detected above. Those can also be avoided if the composition laws employed are being changed according to the Ext and Tor terms in the exact sequence of the universal coefficient theorem. Bulk non-geometric terms are in the opinion of the authors effects due to transitions between different coefficient structures in generalised cohomology. Invariance with respect to that choice amounts to modifications of the composition laws in the bulk leading to effects commonly called "non-geometrical". Such a choice of name is somewhat unfortunate. Bulk quantum corrections involve the replacement of the classical Ryu Takayanagi formula with its quantum extension which takes into account the bulk entanglement. This term itself appears as an extension required to make the theory invariant at changes of coefficients in homology. If the perturbations considered are first order the quantum extension term can be written as an integral over a local quantity. When calculating the constraints on the bulk geometry starting from the quantum Ryu Takayanagi formula such a term produces an expectation value of the stress tensor which can be considered a source for the linearized Einstein's equations. If the CFT states correspond to coherent states in the bulk, the extended entanglement term is zero and hence the bulk doesn't contribute to an entanglement entropy. In general however this is not the case and bulk entanglement plays an important role. The equivalence between bulk relative entropy, seen as the non-local part of the bulk entanglement entropy, and the CFT relative entropy are important in the emergence of locality. The CFT relative entropy for multi-trace states is equivalent to the relative entropy of the corresponding bulk state. 
 In order to produce coherent states of the perturbative bulk fields we use the holographic CFT states defined by 
 \begin{equation}
 \Bracket{\phi_{0}|\Psi_{\lambda}}=\int_{\tau<0}^{\phi(\tau=0)=\phi_{0}}[d\phi(\tau)]e^{-S_{E}-\int_{-\infty}^{0}d\tau \int dx\lambda_{\alpha}(x,\tau)\mathcal{O}_{\alpha}(x,\tau)}
\end{equation}
with $\mathcal{O}_{\alpha}(x,\tau)$ corresponding to the low dimensional operators dual to light fields in the bulk. $\ket{\Psi_{\lambda}}$ is a wave functional which requires the specification of a family of couplings $\{\lambda_{\alpha}\}$. To define excited states of the original theory the sources have to vanish in the $\tau\rightarrow 0$ limit. 
In order to represent squeezed states of the bulk modes or states with entanglement between distant particles in the CFT we may consider multi-trace states of the form 
\begin{equation}
\begin{array}{c}
\Bracket{\phi_{0}|\Psi_{\lambda}}=\int_{\tau<0}^{\phi(\tau=0)=\phi_{0}}[d\phi(\tau)]e^{-S_{E}-S_{\{\lambda^{(i)}\}}},\\
\\
S_{\{\lambda^{(i)}\}}=\sum_{n}\int_{-\infty}^{0}d\tau_{1}...d\tau_{n}\int dx_{1}...dx_{n}\lambda^{(n)}_{\alpha_{1},...,\alpha_{n}}(x_{1},\tau_{1},...,x_{n},\tau_{n})\mathcal{O}_{\alpha_{1}}(x_{1},\tau_{1})...\mathcal{O}_{\alpha_{n}}(x_{n},\tau_{n})
\end{array}
\end{equation}
In the limit in which the bulk theory is free, states with single and non-local double-trace sources give rise to general Gaussian states of the bulk effective theory
\begin{equation}
e^{c_{\alpha,n}a^{\dagger}_{\alpha,n}+c_{\alpha\alpha',nn'}a^{\dagger}_{\alpha,n}a^{\dagger}_{\alpha',n'}}\ket{0}
\end{equation}
which are fully determined by their two-point functions. Multi-trace path integral states correspond to more general excitations of low-energy modes in semi-classical holography. There are however more general situations corresponding to superpositions of coherent states which do not correspond to low energy excitations around a simple gravitational background geometry. A more general structure of bulk entanglement that becomes more relevant to our case of invariance to changes of coefficients in cohomology is the situation in which the bulk connectivity changes. In particular a theory can be defined on a disconnected spacetime with CFTs living on both disconnected pieces. Adding the sources that couple the two separate CFTs leads to 
\begin{equation}
\Bracket{\phi_{1}\phi_{2}|\Psi_{\lambda}}=\frac{1}{Z_{\lambda}}\int [d\phi_{1}][d\phi_{2}]e^{-S_{1}-S_{2}-\int dx_{1}dx_{2}\lambda^{2}(x_{1},x_{2})\mathcal{O}(x_{1})\mathcal{O}(x_{2})}
\end{equation}
we obtain CFTs that are each in a mixed state. This is translated into the fact that the matter on one side is entangled with the matter on the other side. This is similar with the thermofield double state [12] where instead of coupling sources the path integral is defined on a connected spacetime. It so seems that the appearance of sources leads to a connectivity that was not visible before, and that a situation in which a thermofield double state is defined on a connected spacetime is similar with one in which coupling sources are defined on a disconnected spacetime given corrections derivable from the universal coefficient theorem. This is a first example of coefficient changes that lead to a different ability to detect connectivity. 
The replica trick is a method capable of computing entanglement entropies of a subsystem for states defined via Euclidean path integrals in which we can introduce sources. 
Given the mixed state
\begin{equation}
\Bracket{\phi_{-}|\rho|\phi_{+}}=\frac{1}{Z_{\lambda}}\int_{\phi(\tau\rightarrow 0_{-})=\phi_{-}}^{\phi(\tau\rightarrow 0_{+})=\phi_{+}}[D\phi]e^{-S-\int dx_{1}\lambda^{(1)}(x_{1})\mathcal{O}(x_{1})-\int dx_{1}dx_{2}\lambda^{(2)}(x_{1},x_{2})\mathcal{O}(x_{1})\mathcal{O}(x_{2})+...}
\end{equation}
where the integral is over the Euclidean space except for a cut at $\tau=0$. For a spatial subsystem $A$ we can define a density matrix as a partial trace over the complementary subsystem $A^{c}$ of the full density matrix after performing the identification $\phi_{-}=\phi_{+}$ and integrating over $\phi_{\pm}|_{A^{c}}$. The path integral will be then over the Euclidean space minus the region $A$ on the $\tau=0$ slice
\begin{equation}
\Bracket{\phi_{-}^{A}|\rho_{A}|\phi_{+}^{A}}=\frac{1}{Z_{\lambda}}\int_{\phi(\tau\rightarrow 0^{+},x\in A)=\phi_{+}^{A}}^{\phi(\tau\rightarrow 0^{-},x\in A)=\phi_{-}^{A}}[D\phi]e^{-S-\int dx_{1}\lambda^{(1)}(x_{1}) \mathcal{O}(x_{1})-\int dx_{1}dx_{2}\lambda^{(2)}(x_{1},x_{2})\mathcal{O}(x_{1})\mathcal{O}(x_{2})+...}
\end{equation}
For this reduced density matrix we can compute the entanglement entropy as
\begin{equation}
S_{A}=-tr_{A}(\rho_{A} \cdot log \rho_{A})=-\frac{d}{dq}log tr_{A}(\rho^{q}_{A})|_{q=1}
\end{equation}
where we consider an analytic continuation in $q$. The trace then can be evaluated as $tr_{A}(\rho_{A}^{q})=\frac{Z_{q}}{Z^{q}}$ where $Z_{q}$ is the path integral over the replica manifold $M_{q}$ resulting from gluing $q$ copies of the CFT spacetime $M_{1}$ cyclically along the region $A$ keeping the transition of the field $\phi$ from one sheet to the next smooth. The sources are being replicated over each sheet of $M_{q}$
\begin{equation}
Z_{q}=\int_{M_{q}}[D\phi]e^{-S-\int dx_{1}\lambda^{(1)}(x_{1})\mathcal{O}(x_{1})-\int dx_{1}dx_{2}\lambda^{(2)}(x_{1},x_{2})\mathcal{O}(x_{1})\mathcal{O}(x_{2})+...}
\end{equation}
which means the entropy requires the calculation of the Euclidean path integrals $Z_{1}$ and $Z_{q}$ over $M_{1}$ and the replica manifold $M_{q}$.
The perturbative contributions to the partition function at order $N^{2}$ must come from tree diagrams. Entanglement entropies for path integral states with general multi-trace sources are calculated in leading order $N^{2}$ by means of the replica method. We may start with a mixed stated of the form 
\begin{equation}
\Bracket{\phi_{-}|\rho|\phi_{+}}=\frac{1}{Z_{\lambda}}\int_{\phi(\tau\rightarrow 0_{-})}^{\phi(\tau\rightarrow 0_{+})=\phi_{+}}[D\phi]e^{-S=\int dx_{1}\lambda^{(1)}(x_{1})\lambda^{(1)}(x_{1})\mathcal{O}(x_{1})-\int dx_{1}dx_{2}\lambda^{(2)}(x_{1},x_{2})\mathcal{O}(x_{1})\mathcal{O}(x_{2})+...}
\end{equation}
with a cut at $\tau=0$ not covered by the Euclidean space over which we integrate. 
For a holographic CFT perturbed by a nonlocal multi-trace deformation we have a change of the Euclidean action in the form 
\begin{equation}
S_{E}\rightarrow S_{E}+\sum_{n}\int dx_{1}...dx_{n}\lambda^{(n)}(x_{1},...,x_{n})\mathcal{O}(x_{1})...\mathcal{O}(x_{n})=S_{\{\lambda^{(i)}\}}
\end{equation}
The sources $\lambda^{(n)}$ are all of order $N^{2}$. The operators are normalised resulting in all connected $N$-point functions to be of order $(1/N)^{2(n-1)}$. The order $N^{2}$ contributions to $log[\{\lambda^{(i)}\}]$ correspond to fully connected tree graphs. The terminal nodes of any non-vanishing tree diagram must correspond to insertions of a single trace source. While the multi-trace sources affect the partition function at order $N^{2}$ in the presence of single trace sources, multi-trace sources on their own do not affect the partition functions or entanglement entropies at order $N^{2}$. Perturbative contributions to entanglement entropy at order $N^{2}$  vanish when considering the entropy of the whole CFT. The order $N^{2}$ entanglement entropy for a subsystem $A$ computed perturbatively in the sources is always the same as for the complementary subsystem $A^{c}$. This also suggests that for states of the type considered here, the CFT entropy at order $N^{2}$ has a purely geometrical interpretation on the gravity side via the classical RT formula. It has been shown that the entanglement entropy at order $N^{2}$ for states defined by general multi-trace sources is identical to that of another state constructed exclusively from single trace sources where the effective single trace sources are determined from the original sources by means of certain self-consistency equations. The geometrical character of the $O(N^{2})$ entanglement for single trace states is therefore mapped exactly to this more general class of states. 
%%%

Higher $1/N$ corrections are however different. They will fundamentally break the perturbative approach in the sources and hence the bulk interpretation can no longer be assumed to be classical. Using the replica trick, we have seen that the $N^{2}$ contributions emerge from tree diagrams which can be topologically represented on a contractible replica manifold. These will produce contributions to the entanglement entropy that are analytic in the sources. When $1/N$ corrections are included the diagrams will be represented on topologically non trivial manifolds in which loops may wind around non-trivial cycles in the replica manifold. These will produce contributions to the entanglement entropy that are indeed non-analytic in the sources, introducing logarithmic terms in them. The main point of this article is to show that this is a general feature resulting from the requirement of coefficient change invariance induced by the universal coefficient theorem.

It is therefore important to consider the topological properties of the space formed by the ribbon graph diagrams and particularly to see in what cases non-planar diagrams behave like planar diagrams.  One observes that each loop forms a polygon (simplified, it encompasses a 2-simplex) that can be glued together with another index loop in order to form a surface.  The topology of a space can be determined at various homological ”resolutions” with various homological-algebraic tools.  Here ”resolution” refers to the visibility of a certain topological feature.  It has nothing to do with the scale at which one looks and with the analogy to a ”magnifying glass”.  Maybe the best analogy would be a detector sensible to certain topological properties if suitably tuned.  In general, what the physicist wishes is to integrate over such a topological space in order to obtain the answers encoded in the diagrams defined on it. Indeed this is what we require when we calculate our Euclidean path integral and we construct our replica manifold. For this, one needs an integration measure.  This measure is in general sensitive to the (co)homology of the space.  The(co)homologies are invariants of the topological space of a certain accuracy (strength).  They are in general defined in terms of chain complexes.  The chain complexes are defined starting from the q-simplexes $\Delta^{q}$
\begin{equation}
\Delta^{q}=\{(t_{0},t_{1},...,t_{q})\in \mathbb{R}^{q+1}|\sum t_{i}=1,t_{i}\leq 0 \forall i\}
\end{equation}
together with face maps
\begin{equation}
f^{q}_{m}:\Delta^{q-1}\rightarrow \Delta^{q}
\end{equation}
defined as 
\begin{equation}
(t_{0},t_{1},...,t_{q-1})\rightarrow (t_{0},...,t_{m-1},0,t_{m},...,t_{q-1})
\end{equation}
This abstract construction must be mapped into a realistic space $X$. In order to do this a continuous map is required
\begin{equation}
\sigma:\Delta^{q}\rightarrow X
\end{equation}
Considering this, any space can be constructed as a chain
\begin{equation}
\{X\}=\sum_{i=1}^{l}r_{i}\sigma_{i}
\end{equation}
where $\{r_{i}\}$ is the set of coefficients belonging in general to a ring $R$.  The space $X$ as seen via the basis formed from the q-simplexes defined above is denoted $S_{q}(X;R)$.  One defines a boundary map as 
\begin{equation}
\partial : S_{q}(X;R)\rightarrow S_{q-1}(X;R)
\end{equation}
such that 
\begin{equation}
\partial (\sigma)=\sum_{m=0}^{q}(-1)^{m}\sigma\circ f_{m}^{q}
\end{equation}
One can extend the above definition by introducing the covariant functor $S_{*}(-;R)$. This means that given a continuous map
\begin{equation}
f:X\rightarrow Y
\end{equation}
this will induce a homomorphism 
\begin{equation}
f_{*}:S_{*}(X;R)\rightarrow S_{*}(Y;R)
\end{equation}
with the definition
\begin{equation}
f_{*}(\sigma)=f\circ\sigma
\end{equation}
Then, the complex $(S_{*}(X;R),\partial)$ is called the simplicial chain complex of the space $X$ with coefficients in $R$.  The homology of this chain complex with coefficients in $R$ is then 
\begin{equation}
H_{q}(X;R)=\frac{ker \partial}{Im \partial}
\end{equation}
where $ker$ represents the kernel of the considered map and $Im$ represents its image. Hence  the  homology  groups  depend  on  the  coefficient  rings $R$ used  to  define  them.   For  simplicity  one  can  also restrict the rings to groups.  I showed in ref.  [13] following [14] that the universal coefficient theorems can express the  (co)homology  groups  of  a  space  with  a  certain  coefficient  group  in  terms  of  (co)homology  groups  of  the  same space  with  a  different  coefficient  group. I  also  showed  in  [13]  following  [14]  that  some  information  visible  when  a certain group is used becomes invisible when another group is used.  The main idea is that the information about the homotopy class of a function may be accessible when a certain coefficient group is used, while not visible from the (co)homological perspective when another coefficient group is used.  However, the universal coefficient theorem, written as 
\begin{equation}
0\rightarrow Ext(H_{n-1}(C),G)\rightarrow H^{n}(C;G)\rightarrow Hom(H_{n}(C),G)\rightarrow 0
\end{equation}
for cohomology gives us the extra information related to the homotopy classes, just that in this case encoded in the ”homological obstruction” given by the $Ext$ respectively $Tor$ groups.  Here $H_{*}(C)$ is the $*$-th dimension homology of the chain complex $C$ ,$H^{*}(C;G)$ is the $*$-th dimensional cohomology of the chain complex $C$ measured with coefficients in $G$, $Hom(H_{*}(C),G)$ is the group of all homomorphisms from $H_{*}(C)$ to the coefficient group $G$, $H_{*}(C;G)$ is the $*$-th dimensional homology group with coefficients in the group $G$ and the $Ext$ and $Tor$ functors are here the extensions and the torsions of the respective homologies.  These behave as obstructions to the exactness of the short sequence where they would be absent. 

Translated in terms of $\frac{1}{N}$ expansions, this would mean that, when using a certain coefficient group, the integration measure may ”see” the non-planar graphs as planar while the formal differences between the two types can be found only in the form of $Ext$ and $Tor$ groups and modified group operations.  In this way, one can relate theories containing non-planar diagrams, considered hard to solve today, to theories containing only planar diagrams and homological-algebraic corrections to some composition rules.  These corrections will differ for each topological genus they originate from.  This would make many theories exactly solvable if the above mentioned corrections are correctly understood.  In some sense, this amounts, loosely speaking, to a renormalization procedure:  the ”non-solvability” due to non-planar contributions  is  eliminated,  maintaining  the  relevant,  computable  ”non-planar”  contributions  only  in  the  form  of modifications of group laws as specified by the $Ext$ and/or $Tor$ groups. 

In a sense, until now, the lack of strength of topological invariants (their inability to discern some topological spaces)was certainly not seen as a desirable property.  However, several physical and finally natural phenomena can also not make the distinction between some topological spaces defined in terms of chain complexes.  For example, naturally a measure of an integral over a topological space is related more to the cohomology of the space than to its actual mathematically perfectly described shape.  Natural phenomena are also defined in terms of integrations over spaces with certain measures.  This reminds us of the coarse graining in the problems analysed via renormalisation groups. However, in this case it is not the large scale that hides features but the topological invariants and other homological tools that we naturally use.  In this sense I see this idea as a generalisation of ”renormalisation group approaches”. The renormalisation group transformations now become changes in the group structures used in (co)homology.  The regularisation step becomes the identification of the problems originating from the non-planar nature of the corrections and the translation of these into the language of derived functors ($Ext$ and $Tor$).  The standard example of how a function  that  looks  homotopic  to  a  constant  in  the  cohomology  with  a  set  of  coefficients,  is  in  fact  homotopically non-trivial when analysed with another set of coefficients, is presented in what follows. 

If $C$ is a chain complex of free abelian groups, then there are natural short exact sequences of the form 
\begin{equation}
0\rightarrow H_{n}(C)\otimes G\rightarrow H_{n}(C;G)\rightarrow Tor(H_{n-1}(C),G)\rightarrow 0
\end{equation}
$\forall n,G$ and these split. Here $Tor(H_{n-1}(C),G)$ is the torsion group associated to the homology.  In this way homology with arbitrary coefficients can be described in terms of homology with the “universal” coefficient group $\mathbb{Z}$. This result is also valid for cohomology groups.  Moreover, it is a property of algebraic topology independent of the existence of an underlying manifold structure for the spaces or groups on which it may be applied. The following example [4] shows how the choice of the coefficient group can affect the correct identification of the homotopy type of a function. 

Take a Moore space $M(\mathbb{Z}_{m},n)$ obtained from $S^{n}$ by attaching a cell $e^{n+1}$ by a map of degree $m$.  The quotient map $f:X\rightarrow X/S^{n}=S^{n+1}$ induces trivial homomorphisms on the reduced homology with $\mathbb{Z}$ coefficients since the nonzero reduced homology groups of $X$ and $S^{n+1}$ occur in different dimensions.  But with $\mathbb{Z}_{m}$ coefficients the situation changes, as we can see considering the long exact sequence of the pair $(X,S^{n})$, which contains the segment 
\begin{equation}
0=\tilde{H}_{n+1}(S^{n};\mathbb{Z}_{m})\rightarrow \tilde{H}_{n+1}(X;\mathbb{Z}_{m})\xrightarrow{f_{*}}\tilde{H}_{n+1}(X/S^{n};\mathbb{Z}_{m})
\end{equation}
Exactness requires that $f_{*}$ is injective, hence non-zero since $\tilde{H}_{n+1}(X;\mathbb{Z}_{m})$ is $\mathbb{Z}_{m}$, the cellular boundary map 
\begin{equation}
H_{n+1}(X^{n+1},X^{n};\mathbb{Z}_{m})\rightarrow H_{n}(X^{n},X^{n-1};\mathbb{Z}_{m})
\end{equation}
being exactly 
\begin{equation}
\mathbb{Z}_{m}\xrightarrow{m} \mathbb{Z}_{m}
\end{equation}
One can see that a map $f:X\rightarrow Y$ can have induced maps $f_{*}$ that are trivial for homology with $\mathbb{Z}$ coefficients but not so for homology with $\mathbb{Z}_{m}$ coefficients for suitably chosen $m$.  This means that homology with$\mathbb{Z}_{m}$ coefficients can tell us that $f$ is not homotopic to a constant map, information that would remain invisible if one used only $\mathbb{Z}$-coefficients. Let me be more accurate and translate this into notions related to integration.  This has important consequences in the way we calculate the terms in the topological expansion and the calculation of entanglement entropy via the replica trick for higher order corrections in $1/N$ but also in practical calculations of integrals over topologically non-trivial manifolds in general.  In principle the homology groups, $H^{k}(C)$ relate to the shape of the manifold.  The cohomology groups, $H^{k}(C)$ relate to the differential forms defined over the manifold.  Hence, if there is a manifold $M$ characterised by a sequence of homology groups, then, one can define the integral
\begin{equation}
\int_{M}\omega
\end{equation}
characterised by the differentia form $\omega$ and by the manifold $M$. The differential form may encode the integration involved in calculating the diagrams of a genus term. Integration can be seen as the pairing 
\begin{equation}
H_{k}(M,\mathbb{R})\times H^{k}(M,\mathbb{R})\rightarrow \mathbb{R}
\end{equation}
such that 
\begin{equation}
([M],[\omega])\rightarrow \int_{M}\omega
\end{equation}

where this pairing is constructed with real coefficients and this coefficient structure characterises also the measure of integration and implicitly the differential form $\omega$.  Here, $[M]$ represents a class in homology and $[\omega]$ represents a class in cohomology.  The pairing above however is an isomorphism (one-to-one relation) only when this particular choice of coefficients is made.  For other coefficients this pairing may fail to be an isomorphism.  The correction is encoded in a term controlled by the $Ext$ and $Tor$ groups
\begin{equation}
H_{k}(M,\mathbb{G})\times H^{k}(M,\mathbb{G})\rightarrow \mathbb{G}
\end{equation}
where the map becomes 
\begin{equation}
([M],[\omega])\rightarrow \int_{\{M\}}\omega \oplus \int_{Ext(H_{n-1}(C),\mathbb{G})}\eta
\end{equation}
Here,  the first integral is over the surface $\{M\}$ visible when the coefficient structure $\mathbb{G}$ is used and the correction appears as an integral of another differential form over the extension group constructed from the homology with general integer coefficients over a lower dimension.  Here I simply used the universal coefficient theorem in cohomology.  The non-trivial topology however is not visible from the lower dimension hence the simplification. In this way I show that properties defined on some more complex topological objects may be acceptably described on simpler topological objects if controlled changes in the groups used to describe them are being employed. I wish to underline again, not only the fact that this method brings obvious simplifications for the calculation of the topological genus expansion but is also relevant in any field where the integration over topologically non-trivial objects is required. 

With this construction let us return to the calculation of the $1/N$ corrections to the entanglement entropy. As mentioned at the beginning the non-geometric term in the entanglement entropy $\Delta S_{\Sigma}^{bulk}$ originates in the requirement of harmonising the path integral over topologically non-trivial manifolds of the loop diagrams when the coefficient structure is changed such that the structure becomes topologically trivial. 
One may consider as an example the calculation of a black hole radiation entropy. The calculation of this von-Neumann entropy done by Hawking predicted a monotonically growing entropy for the black hole radiation. This calculation is in contrast with the Page curve that takes into account the unitarity of the evaporation process explicitly. When considering a replica trick approach to this calculation one has to integrate over a replica manifold which in certain cases allows for spatially disconnected regions [15,16]. In order to calculate the von Neumann entropy by means of the replica trick we analyse $Tr[\rho^{n}]$ as $n$ copies of the original system. With this, the quantum field theory expands into a theory computed on $n$ copies of the original theory and a set of boundary conditions connecting the various copies cyclically. This cyclic boundary condition can be seen as the insertion of a twist operator in the quantum field theory described over the $n$ copies of the original system. The correlator of twist operator is dual to a partition function of a theory on a topologically non-trivial manifold. The entropy itself can be evaluated by analytical continuation in $n$. This twist operator introducing a series of non-trivial topologies via boundary conditions can be analysed from the perspective of cohomology theory with cyclical coefficients in the following way. If we want to calculate the (co)homology of a circular space the result depends on the type of coefficients. Usually, a topologically non-trivial space with a large circle being uncontractable is detectable by means of homology with integral coefficients. If however the local coefficients gain a non-trivial twist, the homology becomes trivial. Otherwise stated, the twisted homology of a circle with coefficients in $\mathbb{C}$ and non-trivial monodromy vanishes. This implies that a twisted homology of this type completely ignores those parts of the manifold formed by circles along which the monodromy of the coefficient system is non-trivial. That means that the twisted acyclicity of a circle implies that the complement of a tubular neighbourhood of a link will look like a closed manifold because the boundary, as it is fibered to circles, is invisible in the twisted homology. The same remains valid for a collection of pairwise transversal generically immersed closed manifolds of codimension $2$ in arbitrary closed manifolds with the monodromy around each manifold for the coefficients being non-trivial. The twisted homology does not feel the intersection of the submanifolds as a singularity. This result combined with classical results about signatures of manifolds and the relation between twisted homology and homology with constant coefficients makes it possible to analyse a link of codimension two in the same way we analyse a single closed manifold. This is a result proven in ref. [17]. This last part is important for the following reason. The demand of independence of coefficient structures we place on the quantum field theory requires a modification of the entropy leading to terms that encode the Page behaviour of the entropy of the black hole. The tension between the Hawking result and the Page result assuming unitarity of the evaporation process can be linked to non-perturbative effects in the evaluation of the entanglement entropy. These can be linked to the evaluation of the replica trick integral on the $n$-copies manifold in the case in which non-trivial topology becomes manifest. In the $n\rightarrow1$ limit of the replica manifold the description can be made in two ways: either we use acyclic cohomology with coefficients that make the topological features invisible but we must introduce $Ext$ resp. $Tor$ corrections to the integral, or we calculate the integral over a general non-twisted cohomology but then we have to take into account the topological non-triviality of the links involved. The later method has been used in [11,15,16] with promising results. In this approach the twisting operators are being represented via $Tor$ contributions to the universal coefficient theorem in homology. This will result in a modification of the action resulting from the $Tor$ contribution to the integral. This modification will account for the generalised entropy and leads to the quantum extremal surface prescription. 
The universal coefficient theorem can therefore become important in understanding the entanglement entropy of the Hawking radiation. Indeed, the appearance of a horizon separates a region of spacetime into two disconnected domains that cannot classically communicate. That is equivalent to a change in topology. If one accepts the idea of invariance to topology changes as mentioned in ref. [13] one has to reformulate the theory in a topologically invariant way, and from a universal coefficient theorem point of view this would amount to describing the theory starting with the general UCT exact sequence for general types of coefficients. This is of course quite complicated as the number of different types of coefficients is technically unlimited. There are however certain coefficients that are important for the situation in which we want to take into account short range entanglement between outgoing and incoming modes on the two sides of the black hole horizon. Sufficient to say that in order to make the path integral expression on which the calculation of the entanglement entropy via the replica trick relies, topology independent, one has to introduce corrections resulting in additional terms. The details starting from basic principles have been shown in ref. [13]. Calculation of the path integral required for the estimation of the results of the replica trick must involve summation over all topologies connecting the various replicas, involving therefore spacetime wormholes. These wormholes appear to connect the copies of the original black hole. Considering the end of world brane of ref. [18] and an orthogonal basis for internal end-of-world brane states, depending on the dimension of this basis there are two dominant topological regimes. One which arises in the small limit where the topologies seem to be disconnected, and one in the large limit where the topologies are connected. The replica symmetry group changes accordingly, giving for the connected topology $\mathbb{Z}_{n}$ and after dividing with this symmetry the black hole geometry is recovered with the observation that a conical singularity appears at the fixed point of the replica symmetry. When we move to a single replica the singularity vanishes. Analysing this on the page curve we observe that at early times the disconnected replica topologies are the most important. There exist also quantum extremal surfaces just inside the event horizon. After the Page time this contribution becomes dominant and the topology becomes fully connected. Hence, the evolution of the Page curve follows a transition from a dominant disconnected topology to one that is fully connected. This requires for a topology independent description which has been performed in [13] with the observation that the correction terms arising in the universal coefficient theorem must be taken into account in order to correct the replica trick path integral. Therefore let us apply the universal coefficient theorem to the replica manifold transition from a disconnected topology to a connected topology in the context of the Page curve. A similar reasoning will apply for the topological series expansion in string theory. 
In a Jackiw Teitelboim theory of gravity coupled to a non-gravitating two-dimensional conformal field theory encoding the radiation away from the black hole the Page curve is corrected by means of the entanglement entropy modified by the so called island formula
\begin{equation}
S_{A}=min_{B} ext[\frac{\mathcal{A}(\partial B)}{4G_{N}}+S_{AB}^{eff}]
\end{equation}
The island is referred to as $B$ in the gravitating region and $\mathcal{A}(\partial B)$ is the area of the boundary of the island, while $S_{AB}^{eff}$ is the effective field theory entanglement entropy of the quanta in the union regions of $A$ and $B$. By employing this island, the entropy growth of the Hawking radiation terminates at the Page time. The island formula has also been derived by calculating the radiation entropy via the replica trick employing Euclidean wormholes between the included replicas. Another calculation in ref. [18] shows that the same result can be reproduced without considering an island and that the island considered before represents the region between the original horizon and the end-of-world brane used in reference [18]. The "islands" in [18] are seen as regions of space disconnected from the AdS boundary but with the possibility imbedded in them to be reconstructed thanks to their entanglement with the radiation. When the inception black hole of [18] is being purified by means of a two boundary wormhole there is a topology change that is seen as the emergence of islands on the spatial sliced of the glued geometry. This is particularly amenable to a universal coefficient theorem approach in which the two different coefficients may miss or detect a topology and the underlying integral can be constructed such that topological invariance must be preserved. Indeed, if we use the replica trick with the euclidean wormhole extension the resulting path integral is modified by the non-trivial topology. If however the integration is being done through surfaces described by cohomology groups with cyclic coefficients, the non-trivial topologies disappear and the emerging integral corrections amount to the same terms required to provide the proper Page behaviour of the entanglement entropy. I showed this extensively in [13] and partially in [14]. The link between twisted coefficients and the relativity of the circles I also showed in [19]. Indeed, the calculation of ref. [19] showed clearly that the change in coefficient structures performs a transition from two independent circles to the wedge sum between the two circles tangent at a common point, then to a single circle and finally to a simple point.

\section{Conclusions}
Recently, the calculation of entanglement entropy in a situation with emerging non-trivial topology resulted in a recovery of the Page curve. Similar calculations have been performed for the Island model and for the end-of-world model. In this article I use a result I have been discussing several years ago in the context of the Hawking radiation to show that by its application and the demand for topological invariance of the theory the Page curve is being equally recovered, reducing the discussion of how to interpret the islands to a discussion about an arbitrary choice of coefficients in the cohomology encoding the visible topology of our replica manifold. The universal coefficient theorem links the situation in which the coefficients permit the detection of the non-trivial topology to that in which the non-trivial topology is invisible while island terms need to be added to adjust for the non-exactness of the short sequence in the universal coefficient theorem. The procedure amounts to a recalculation of the form of the path integral via Ext/Tor terms that encode the additional entanglement entropy and the correction resulting in the Page curve. The multipartite situation has also been discussed in ref. [20]. 

\end{document}